# Predicting Sediment and Nutrient Concentrations in Rivers Using High Frequency Water Quality Surrogates

Catherine Leigh[1,2,3], Sevvandi Kandanaarachchi[1,4], James M. McGree[1,3], Rob J. Hyndman[1,4], Omar Alsibai[1,2,3], Kerrie Mengersen[1,3], Erin E. Peterson[1,2,3]

[1]ARC Centre of Excellence for Mathematical & Statistical Frontiers (ACEMS), Australia.

[2]Institute for Future Environments, Queensland University of Technology, Brisbane, Queensland, Australia.

[3]School of Mathematical Sciences, Science and Engineering Faculty, Queensland University of Technology, Brisbane, Queensland, Australia.

[4]Department of Econometrics and Business Statistics, Monash University, Clayton, Victoria, Australia.

Corresponding author: Catherine Leigh (catherine.leigh@qut.edu.au)

**Key Points:**

- High frequency prediction of sediments and nutrients from low-cost surrogates will improve water resource monitoring and management

- Dissolved nutrient prediction from surrogates may involve substantial uncertainty due to complexities of ecosystem and environmental drivers

- Turbidity is a useful surrogate of sediments, measured as total suspended solids, in rivers that flow into the Great Barrier Reef lagoon

## Abstract

A particular focus of water-quality monitoring is the concentrations of sediments and nutrients in rivers, constituents that can smother biota and cause eutrophication. However, the physical and economic constraints of manual sampling prohibit data collection at the frequency required to capture adequately the variation in concentrations through time. Here, we developed models to predict total suspended solids (TSS) and oxidized nitrogen (NOx) concentrations based on high-frequency time series of turbidity, conductivity and river level data from low-cost *in situ* sensors in rivers flowing into the Great Barrier Reef lagoon. We fit generalized least squares linear mixed effects models with a continuous first-order autoregressive correlation to data collected traditionally by manual sampling for subsequent analysis in the laboratory, then used these models to predict TSS or NOx from *in situ* sensor water-quality surrogate data, at two freshwater sites and one estuarine site. These models accounted for both temporal autocorrelation and unevenly time-spaced observations in the data. Turbidity proved a useful surrogate of TSS, with high predictive ability at both freshwater and estuarine sites. NOx models had much poorer fits, even when additional covariates of conductivity and river level were included along with turbidity. Furthermore, the relative influence of covariates in the NOx models was not consistent across sites. Our findings likely reflect the complexity of dissolved nutrient dynamics in rivers, which are influenced by multiple and interacting factors including physical, chemical and biological processes, and the need for greater and better incorporation of spatial and temporal components within models.

## 1 Introduction

Water-quality monitoring of streams and rivers is conducted in many regions across the globe, with the goal of protecting human health and the environment. Of particular concern is the concentration of sediments and nutrients in water flowing through rivers, given the potential detrimental effects these constituents have on river health, water treatment, and receiving systems downstream, including marine ecosystems (Brodie et al., 2012; Leigh et al., 2013). In highly seasonal, event-driven river systems, such as those in the tropics, high-magnitude flows during the wet season can transport large quantities of sediments and nutrients from the land downstream in relatively short time frames (O'Brien et al., 2016). However, the rapidity of change in sediment and nutrient concentrations during high-flow events poses challenges for water-quality monitoring based on the manual collection of water samples for subsequent laboratory analysis of constituents. For instance, high flows may preclude safety conditions required for manual sampling, and sample collection at the frequency required to capture change in concentrations may not always be physically or economically practical.

A potential solution is to use relatively low-cost *in situ* sensors automated to measure and record water quality at high frequencies (e.g. every hour), creating multi-parameter time series. These sensors can measure water-quality parameters such as turbidity and conductivity, which have the potential to act as surrogates of sediments and nutrients; constituents that would otherwise need to be measured in the laboratory (Horsburgh et al., 2010; Jones et al., 2011). These low-cost *in situ* sensors have the potential to complement or indeed circumvent the need for manual sampling and laboratory analysis, whilst also providing monitoring data at the frequencies required to capture the full range of water-quality conditions occurring in rivers.

However, models are needed that can effectively predict sediment and nutrient concentrations from the surrogate water-quality time series before the sensor data can be used for such purposes.

A wide range of modelling techniques with the potential to do this exists. For example, Diamantopoulou et al. (2005) used Artificial Neural Networks to predict nitrate from multiple parameters including concentrations of other nutrients, and West and Scott (2016) used standard major axis regression to determine the correlation between turbidity and total suspended solids (TSS; i.e. suspended particles > 2 microns in size). Simple and multiple linear regression, however, appear to be the most commonly used methods for modelling sediment and nutrient concentrations. These methods have been used, for example, to estimate and/or predict suspended solids or suspended sediments from turbidity (Horsburgh et al., 2010; Smackman Jones et al., 2011; TSS, Ruzycki et al., 2014; Skarbøvik and Roseth, 2015; Nasrabadi et al., 2016; O'Brien et al., 2017; Stutter et al., 2017), total nitrogen from turbidity (O'Brien et al., 2017) or turbidity and conductivity combined (Stutter et al., 2017), and total, particulate or dissolved forms of phosphorus from turbidity (Jones et al., 2011; Horsburgh et al., 2010; Ruzycki et al., 2014; Viviano et al., 2014; Skarbøvik and Roseth, 2015; O'Brien et al., 2017, Stutter et al., 2017). However, such models have been criticized for failing to account for temporal autocorrelation in the underlying data, an inherent property of water-quality time series (e.g. Slates et al., 2014). Linear mixed effects models, such as those implemented by Lessels and Bishop (2013) and Slaets et al. (2014) to predict to sediments and nutrients from turbidity, are considered more appropriate precisely because they can be formulated to account for within-group correlation and/or heteroscedasticity through the incorporation of random effects and/or specific variance-covariance structures.

Here, our key objective was to predict concentrations of sediments (i.e. total suspended solids; TSS) and oxidized nitrogen (i.e. nitrite + nitrate; NOx) based on high frequency water-quality data measured by *in situ* sensors in rivers flowing into the Great Barrier Reef lagoon, Australia (Figure 1). To our knowledge, we are the first to predict NOx using models fit to water-quality time series that explicitly account for temporal autocorrelation in data from both estuarine and freshwater sites. Lessels and Bishop (2013) and Slaets et al. (2014), for example, estimated the concentrations of total and particulate nutrients, respectively, not the concentrations of bioavailable, dissolved inorganic nutrients such as NOx. Our main aims were to (i) develop models that include observations of surrogate water-quality parameters as covariates to predict TSS and NOx from *in situ* sensor measurements, and (ii) provide recommendations about the most effective low-cost water-quality surrogates and the potential generalizability of the models across sites.

## 2 Materials and Methods

### 2.1 Study region and sites

Our three study sites are located in rivers that flow into the Great Barrier Reef lagoon along the northeast coast of tropical Australia in Queensland (Figure 1). Two of the sites (Sandy Creek and Pioneer River) lie within the Mackay Whitsunday region and the third (Mulgrave River) lies within the Wet Tropics region. These two regions are characterized by seasonal climate, with higher rainfall and air temperatures in the 'wet' season and lower rainfall and air temperatures in the 'dry' season. Although there is inter-annual seasonal variation in climate and river flow in both regions, the wet season typically occurs from December to April in the Mackay Whitsunday region, and from November to April in the more northern Wet Tropics region (Brodie, 2004; Wang et al., 2011; McInnes et al., 2015). The wet season is typically associated with tropical cyclones, monsoonal rainfall and associated event flows in rivers, and the dry season with low or zero surface flows.

Pioneer River rises in the forested uplands of the Great Dividing Range in north Queensland (Brodie 2004). Many of its upper reaches lie within National or State Parks, whilst land use in the mid and lower reaches is dominated by sugarcane farming. Sandy Creek is a low-lying coastal-plain stream south of the Pioneer River, where the dominant land use is also sugarcane farming. The Mulgrave River in the Wet Tropics World Heritage Area, rises like the Pioneer River in forested National Park uplands of the Great Dividing Range and flows through mostly cleared alluvial floodplains in its lower reaches (Rayner et al., 2007). The Pioneer River and Sandy Creek sites (PR and SC) are located within freshwater reaches, and the Mulgrave River site (MR) is located in an estuarine reach. The monitored catchment area of each site is 1466 km$^2$ (PR), 326 km$^2$ (SC) and 789 km$^2$ (MR). We chose these sites because they had comprehensive water-quality datasets available containing both laboratory-measured sediment and nutrient concentrations, as well as high frequency *in situ* water-quality data from multiple sensors. We selected an estuarine site in addition to the two freshwater sites to investigate the generalizability and broaden the potential application of the models.

2.1 Laboratory and *in situ* sensor data

The Queensland Department of Environment and Science (DES) has installed an *in situ* automated water-quality sensor (YSI EXO2 Sonde attached with an EXO Turbidity Smart Sensor 599101-01 and EXO Conductivity & Temperature Smart Sensor 599870) at each of the three study sites. Sensors are housed in a flow cell in water-quality monitoring stations on riverbanks; water is pumped at regular intervals from the river to the flow cell, approximately every hour and sometimes more frequently during event flows, for the sensors to measure and record turbidity (NTU) and electrical conductivity at 25 °C (conductivity; µS/cm). Pressure-induction sensors record river level (i.e. height in meters from the riverbed to the water surface; level, m) every 10 minutes. Linear interpolation of the ten-minute data provide time-matched observations of level for each observation of turbidity and conductivity. The sensor data were quality assured by DES prior to analysis (Leigh et al., in review) to remove anomalies due to technical errors, which resulted in periods of missing data.

DES manually collect grab-samples of water approximately monthly, and more frequently during event flows in the wet season when safety permits, from each site as part of their Great Barrier Reef Catchment Loads Monitoring Program (Huggins et al., 2017), creating unequally spaced observations of water quality through time. The goal of the program is to track long-term trends in the quality of water entering the Great Barrier Reef lagoon from adjacent catchments, as part of the Paddock to Reef program (Carroll et al., 2012). Collection, storage and transport of grab-samples is conducted under strict quality control and assurance procedures (Standards Australia 1998a, 1998b; DES, 2018). Samples are analyzed in the National Association of Testing Authorities credited Science Division Chemistry Centre laboratories (Dutton Park, Queensland) for turbidity, conductivity and concentrations of TSS (mg/L) and NOx (mg/L) following standard methods (APHA-AWWA-WEF, 2005). DES also record river level on most occasions when grab-samples are collected.

Turbidity is a visual property of water indicative of its clarity (or lack thereof) due to suspended abiotic and biotic particles, which absorb and scatter light. As a result, turbidity tends to increase during high-flow events in rivers, when waters often contain high concentrations of particles (e.g. from runoff-derived soil erosion), and is one of the reasons why turbidity may be a suitable surrogate for TSS (i.e. sediments and other particulates, including nutrients). Turbidity can also increase during low flow phases, due to the resultant concentration of suspended particles, or when high concentrations of microalgae reduce water clarity. River turbidity can change rapidly during flow events, as can conductivity, which reflects the ability of water to pass an electric current. Conductivity is determined by the concentration of ions in the water, including bioavailable nutrients such as NOx. New inputs of fresh water will typically, and quickly, decrease conductivity in rivers as waters dilute. By contrast, conductivity tends to increase during periods of low flow and when water levels decline. Inputs of saline water from groundwater inputs or tidal influence will also increase the conductivity of surface waters.

Turbidity, conductivity, TSS and NOx data measured in the laboratory were available from January 2016 to June 2017 at SC and MR and from January 2016 to October 2017 at PR. Turbidity, conductivity and level data measured and recorded *in situ* by automated sensors were available from March 2017 to March 2018 at all three sites (Table 1).

2.3 Statistical analysis

We fit generalized least squares linear mixed effects models with a continuous first-order autoregressive correlation (AR1) structure (Pinheiro and Bates, 2004) to the laboratory data to predict TSS or NOx (the response variable) from surrogate water-quality variables (the covariates). The models are of the form:

$$y_t = X_t\beta + \varepsilon_t$$

where *y* is an *n*-dimensional vector of TSS or NOx collected at time *t*, *n* is the number of observations, *X* is an *n* × *p* design matrix of *p* covariates, *β* is a *p*-dimensional vector of estimated regression coefficients, and *ε* is an *n*-dimensional vector of zero-mean, normally distributed errors with covariance matrix $\sigma^2\Lambda$. The covariances are defined by a continuous AR(1) structure, such that $Corr(\varepsilon_i, \varepsilon_j) = \phi^{(t_i - t_j)}$, where $t_k$ is the time of the *k*th and $\phi$ is the parameter of the AR(1) process, which can range between 0 and 1. This continuous AR(1) structure was included to account for both the temporal (serial) correlation and unequal temporal spacing present in the laboratory time-series data.

We selected potential covariates for the TSS and NOx models based on plausible mechanisms that could cause changes in TSS or NOx, evidence from the literature, exploratory data analysis and the availability of covariates within the laboratory dataset (following Isaacs et al. 2017). For the TSS model, we included three covariates plus their interactions: turbidity measured in the laboratory, a categorical variable, T15, representing the 'level' of turbidity (low, < 15 NTU; high, ≥ 15 NTU), and a categorical variable for the site (MR, PR, SC). Site was also included in the correlation structure as a grouping variable, to account appropriately for the temporal correlation in the time series site by site. We included T15 because exploratory data analysis indicated (i) a strong and similar positive relationship between turbidity and TSS at all sites (Figure S1), as expected given the physical properties of these variables and the processes underlying water quality dynamics in rivers (Wetzel 2001; Boulton et al. 2014), and (ii) that at lower levels of turbidity, the relationship with TSS appeared to differ from that at higher levels, particularly at the freshwater sites PR and SC (Figure S1). The cut-off value of 15 NTU is the water-quality guideline value for turbidity in freshwater streams and rivers in northern Australia (ANZECC/ARMCANZ, 2000) and matched closely the point of observable change in the bivariate relationship between turbidity and TSS (Figure S1).

We visually examined the relationships between NOx, conductivity, turbidity and level (as measured in the laboratory or on-site). These relationships were not as strong as they were for turbidity and TSS, particularly at MR and PR, and appeared to differ substantially from site to site (Figures S2-4). We therefore fit separate NOx models for each site, including conductivity, turbidity, level and each of the two-way interactions involving level as potential covariates. We chose to do this because each of the potential covariates could theoretically contribute to, or help explain, the concentrations of nutrients in river water (Wetzel, 2001; Boulton et al., 2014). We also included a grouping variable based on river level in the correlation structure for the NOx models to account for the possibility that temporal correlation of the model residuals differed depending on river level. During high-flow events, for example, concentrations of NOx are expected to vary more considerably through time than during more stable flow periods (e.g. Duncan et al., 2017). We defined low level in three different ways for each site (< first quartile, < median, and < third quartile), testing each in the correlation structure in turn at each site to determine and select the preferred model for each site, as described below.

Based on exploratory analyses, we made the decision to fit a single model to all TSS data and separate NOx models for each site. As such, we used different model selection procedures to select the final TSS and NOx models. For TSS, we used a backwards stepwise model selection procedure and maximum likelihood for parameter estimation so that we could use the Akaike information criterion (AIC; Akaike, 1974) to identify the combination of covariates with the most support in the data. Maximum likelihood may produce biased estimates of the correlation parameter (Cheang and Reinsel, 2000). Therefore, we refit this model using restricted maximum likelihood (REML) for parameter estimation before implementing a cross-validation procedure, based on the blocking method of Roberts et al. (2017) for temporally correlated data. We used the results of the cross validation to calculate the cross-validated root mean square error and 95% prediction coverage value (*cvRMSE* and *cvPC*, respectively). This involved dividing the time series from each site into five blocks of chronologically ordered observations (5 blocks x 3 sites = 15 blocks) and then implementing a leave-one-block-out fitting cross validation. We also calculated a cross-validated r-squared (*cvR²*) statistic for the final TSS model as:

$$\left[1 - \left(\frac{cvRMSE}{variance(y)}\right)^2\right]$$

Although it was not deemed appropriate to fit a single model to the pooled NOx data, our aim was to produce as generalizable a model as possible. Thus, the goal of the model selection procedure was to identify a composite model that contained the most important covariates for each site. We did this in three steps. First we implemented a backwards-stepwise model-selection procedure three times for each site to identify the best set of covariates for each model, by site given an AR(1) grouping structure based on the above three definitions of low level. The model with the lowest AIC value at each site was deemed the model with the best subset of covariates. Second, we refit the best set of models from the previous step (3 sites x 3 AR(1) structures = 9 models) using REML and generated the *cvRMSE* for each. For each site, the model with the lowest *cvRMSE* was deemed the single 'best' model in terms of predictive performance. Finally, we combined the covariates found in these three 'best' models to create a final composite NOx model. We refit this final model, using REML, to the data from each site separately and generated the *cvRMSE*, *cvPC*, and *cvR²* for each.

With the turbidity, conductivity and level data from the *in situ* sensors, we then used the final TSS model and three final NOx models to make predictions of TSS and NOx, respectively. The predictions and associated estimates of uncertainty were generated using an infinite-horizon forecast (Hyndman and Athanasopoulos, 2018) for the AR(1) structure because no lagged observations were available for the sensor data. This ensured that the 95% prediction intervals had the correct nominal coverage. We performed all statistical analyses in R and the nlme package was used to implement the linear mixed effects models (Pinheiro et al., 2017). Turbidity, conductivity, TSS, NOx and river level were all $\log_{10}$-transformed prior to analysis. Predictions and prediction intervals from the models were then back-transformed for graphical visualization and interpretation.

## 3 Results

### 3.1 Laboratory and *in situ* sensor data comparisons

We visually compared the laboratory-measured conductivity, turbidity and level data respectively with the conductivity, turbidity, and level data from the *in situ* sensors at each site and found the laboratory and sensor data exhibited similar patterns over time (Figures 2-4). The one exception out of the nine comparisons was conductivity at MR, where the laboratory data covered a much lower range of values than did the *in situ* sensor data (30 - 2900 vs 0.58 - 396 µS/cm; Figure 2, Table 1).

### 3.2 TSS and NOx models

The final TSS model explained just over 90% of the variation in TSS (Table 2). The model covariates included turbidity, site and T15, as well as interactions between site and turbidity and between T15 and turbidity (Table 3, Figure 5, Table S1 and Figures S5-7). The relatively high value of the autocorrelation parameter for the model ($\phi = 0.87$) indicated there was temporal autocorrelation present in the data, as expected. All terms in the model were statistically significant ($p < 0.01$; Table 3).

NOx models that included a grouping structure based on the medial level value had the best predictive ability at each site, based on the *cvRMSE*, but the measures of predictive performance of the final, composite NOx model were not as high as those of the TSS model (Table 2). The final composite NOx model included turbidity as a covariate, along with conductivity and level, as well as the interactions between level and conductivity and level and turbidity (Table 4, Figure 6, Table S2 and Figures S8-10). However, the statistical significance of each of these covariates depended on site (Table 4). For MR, just over 25% of the variation in NOx was explained (Table 2) and conductivity was the only statistically significant covariate ($p < 0.0001$; Table 4). For PR, all covariates were statistically significant ($p < 0.05$; Table 4) except turbidity, but the model had the smallest *cvR²* of all those fitted (just over 17%; Table 2). For SC, conductivity, turbidity and the interaction between conductivity and level had significant effects as covariates ($p < 0.05$; Table 4) and the model explained just over 35% of the variation in NOx (Table 2). Correlation parameters for each of the final NOx models were indicative of the presence of temporal autocorrelation, as expected ($\phi > 0.86$; Table 2).

3.3 TSS and NOx predictions

Predictions using the final TSS model and the turbidity data recorded by the *in situ* sensors indicated that the prediction intervals tended to be wider during events than during non-event periods, for all sites (Figures 7-9; Table S3). The highest sensor TSS value predicted at MR (888 mg/L) was well above that of the highest laboratory-measured TSS value at MR (221 mg/L), despite all laboratory TSS values falling within the 95% prediction interval for the sensor predictions for each site (Table 1, Table S3). The 888 mg/L prediction, however, was associated with the highest sensor-measured turbidity value at MR (396 NTU in October 2017), which was similarly well above that of the highest laboratory turbidity value at MR (143 NTU in January 2017; Table 1, Figure 2).

The wide prediction intervals determined using the *in situ* sensor conductivity, turbidity, and level data were related to the poorer fits and weaker predictive capacity of the NOx models, relative to the TSS model (Table S4, Figures 10-12). This was particularly noticeable for the model applied to MR data. For example, the maximum predicted value of sensor NOx had a prediction interval with a range of more than 30 000 mg/L, which is an unrealistic concentration in itself and well above that observed in the laboratory data (1.0 mg/L at PR; Table 1).

**4 Discussion**

Knowing the concentrations of sediments and nutrients in rivers, and how they change through time, is a key focus of water-quality monitoring. This knowledge helps inform effective protection and management of our land, waterways and oceans, including World Heritage Areas such as the Great Barrier Reef (Schaffelke et al. 2012; Humanes et al., 2017). However, traditional methods of monitoring these constituents rely on discrete manual sampling of water followed by laboratory measurement of concentrations, which is both time consuming and costly. In addition, the relatively low sampling frequency increases the chances of missing water-quality events. The spatial sparsity of concentration measurements from manual sampling is also problematic. For example, the Great Barrier Reef lagoon stretches over 3000 km of coastline, while data used currently to validate estimates of sediments and nutrients flowing to river mouths are generated through a monitoring program that targets just 43 sites. This provides limited knowledge of sediments and nutrient concentrations in both space and time, and emphasizes the importance of developing predictive models for these concentrations based on water-quality surrogates measured at high frequency using low-cost *in situ* sensors.

Studies have shown that turbidity is a useful surrogate of sediments in rivers, particularly when models account for temporal correlation in the data (Lessels and Bishop 2013). This was also the case in our study, where the predictive ability of TSS models based on turbidity was relatively high at both freshwater and estuarine sites. Although the results of some studies suggest that turbidity also holds promise as a surrogate of nutrients in rivers, to our knowledge this has only been demonstrated for total or particulate forms at freshwater sites (e.g. Slaets et al. 2014), possibly because the relationship between turbidity and dissolved forms, such as NOx, is not as strong. Indeed, in contrast to our TSS models, the NOx models had much poorer fits, even when temporal correlation was accounted for and additional covariates of conductivity and river level were included along with turbidity. Furthermore, the relative importance of covariates in the models was not consistent across sites. These results likely reflect the complexity of dissolved nutrient dynamics in rivers, which are influenced by multiple and interacting factors including physical, chemical and biological processes (Wetzel 2001; Boulton et al. 2014). For example, different timings and applications of fertilizers to agricultural land in the watersheds, as well as different spatial configurations and types of agricultural land (e.g. livestock grazing versus sugarcane cropland), may all differentially influence the concentration of dissolved nutrients at each site and through time (Hunter and Walton, 2008; Bainbridge et al., 2009). Our goal in this study was to develop models that were generalizable across sites and so we did not include site-specific land use. However, additional covariates such as percent agricultural land use or flow-weighted land use in the watershed (Peterson et al. 2011) may improve model fits and resultant predictions.

The poorer fits of the NOx models may also be due in part to the limited range of values covered by some of the surrogate covariates used in the models. For example, at the estuarine site, MR, conductivity regularly fluctuates between 10 and 50 000 µS/cm, due to tidal influence. However, the highest observation of conductivity used to build the NOx model for MR was 2900 µS/cm. Applying the *in situ* sensor data from MR, for which the highest conductivity observation was 48 453 µS/cm, to this model thus led to overly high and unrealistic predictions of NOx concentrations. We therefore recommend that predictive models be built using data that cover the range of values expected in the water quality surrogates at the study sites of interest, whenever possible. A broader coverage of data values may also serve to improve the generalizability of models from one site to another. However, separating data from freshwater and estuarine sites may be necessary (as we did for the NOx models here) when predicting water quality variables from surrogates that vary substantially from one environment to the other in their natural values, such as conductivity. We further suggest that pooling data from very differently characterized freshwater sites (e.g. those heavily influenced by ground water inputs versus run-off dominated sites) into a single model may also lead to poor model fits, unless an enlarged model is used that accounts carefully for this site to site variation.

As anticipated, prediction intervals for both TSS and NOx were wider during events, when the predicted concentrations of these constituents increased substantially as turbidity, conductivity and/or river level increased. This has implications for model application; for instance, if these models were used to predict sediment and nutrient concentrations during high flow and high turbidity events, as would be desirable of most water-quality monitoring programs, users would need to be aware that the uncertainty around those predictions may be quite high, especially at the upper end of the prediction interval. These differences in uncertainty are important because they provide managers with information about where and when they can be most/least confident in the predictions (Peterson and Urquhart 2006) so that they can effectively prioritise management actions (Yoccoz et al., 2001). Moreover, collection of surrogate data at different time stamps may limit the number of observations that can be used in models when multiple covariates are involved, and/or may inflate the uncertainty in predictions if, for example, interpolation is used to match the timestamps across observations. Therefore, we recommend that observations for different water-quality surrogates measured by *in situ* sensors be collected at the same time when possible and measures of uncertainty (e.g. prediction intervals) be included routinely in water-quality reporting.

Significant investments are being made to improve management practices, with the goal of reducing the amounts of sediments and nutrients entering Great Barrier Reef waterways (Carroll et al., 2012). However, measuring the downstream impacts of these investments is challenging because current water-quality monitoring relies on a relatively small number of sites at, or near river mouths. This makes pinpointing where the largest sources of sediment and nutrient inputs are within a watershed difficult. The ability to predict TSS and NOx from low-cost *in situ* sensors provides an opportunity to deploy a network of sensors throughout watersheds, which has numerous benefits for management. Firstly, the number of water-quality monitoring sites would increase significantly. Secondly, as the amount of data increases, the opportunity to develop near-real time statistical models for TSS and NOx increases, which could then be used to create dynamic predictive maps of sediment and nutrient inputs throughout entire watersheds. This would provide managers with greater situational awareness of where and when water-quality targets are being breached and would allow prioritization of land management actions in space and time to further reduce land-based impacts on the Great Barrier Reef lagoon.

Our study has highlighted several directions for future research; in particular, we see the need for greater and better incorporation of spatial and temporal components within predictive models. Spatial statistical models for stream networks that capture the spatial dependency in stream data (Peterson and Ver Hoef, 2010) could be used to apply the models developed herein across entire Great Barrier Reef watersheds. If a network of low-cost *in situ* sensors were to be deployed throughout a catchment, it would likely be beneficial to extend these spatial methods into time (i.e. spatio-temporal models; Cressie and Wikle, 2011), for which the need clearly still exists (Peterson et al., 2013; Isaak et al. 2017). Furthermore, because *in situ* sensors are prone to drift and other technical issues, the data they produce are not free from anomalies (i.e. errors). Yet the streaming, near-real time nature of the data produced from these sensors renders manual detection and correction of anomalies for quality control and assurance infeasible (Horsburgh et al., 2015). The extension of both automated anomaly detection methods, developed for high frequency water-quality data (Leigh et al., in review) into space and time on river networks, in combination with spatio-temporal models that predict sediment and nutrient concentrations from these data, could thus revolutionize the way water quality is monitored and managed, whilst also increasing scientific understanding of the spatio-temporal dynamics of sediment and nutrient processes.

**Acknowledgments and Data**

Funding for this project was provided by the Queensland Department of Environment and Science (DES) and the ARC Centre of Excellence for Mathematical and Statistical Frontiers (ACEMS). The authors would like to acknowledge the Queensland Department of Environment and Science, Great Barrier Reef Catchment Loads Monitoring Program for the data and the staff from Water Quality and Investigations for their input. We thank Grace Heron for producing the map in Figure 1. A repository of the laboratory and *in situ* sensor data used in this paper are provided in the Supporting Information.

**Figures**

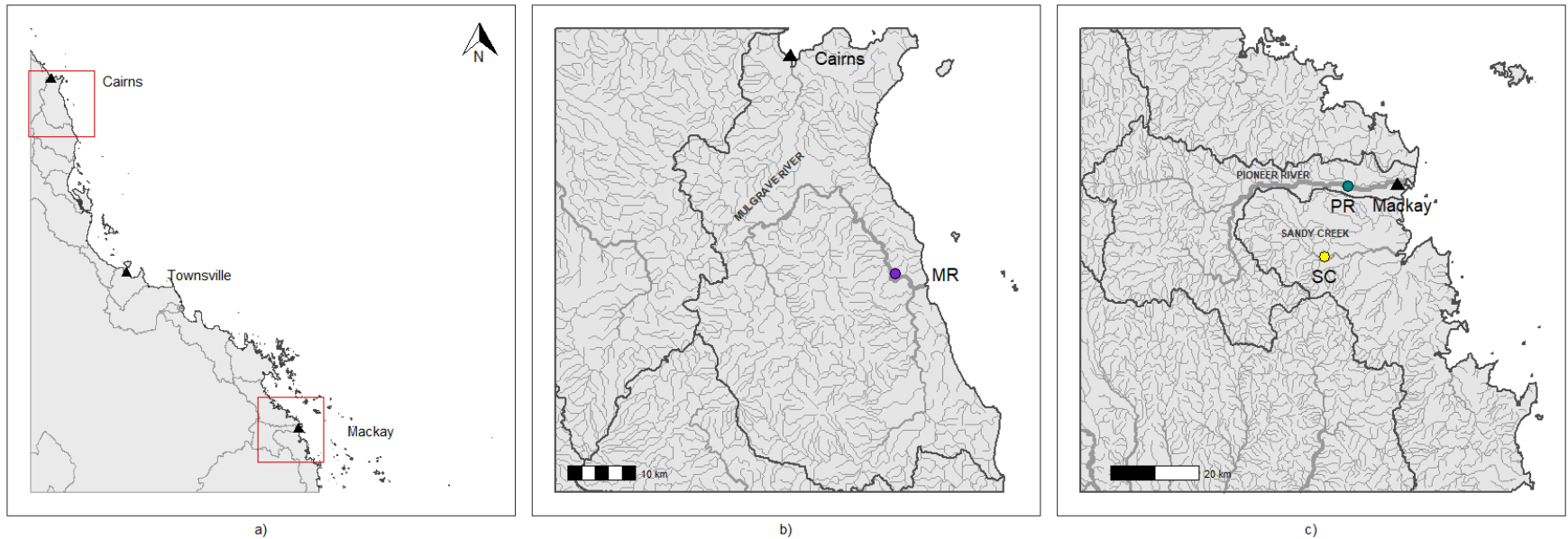

**Figure 1.** The study region in (a) north tropical Queensland, Australia, showing our study sites (closed circles: MR, purple; PR, green; SC, yellow), rivers and watershed boundaries within (b) the Wet Tropics and (c) Mackay Whitsunday regions (PR and SC). Closed triangles show the major towns of Cairns, Townsville and Mackay.

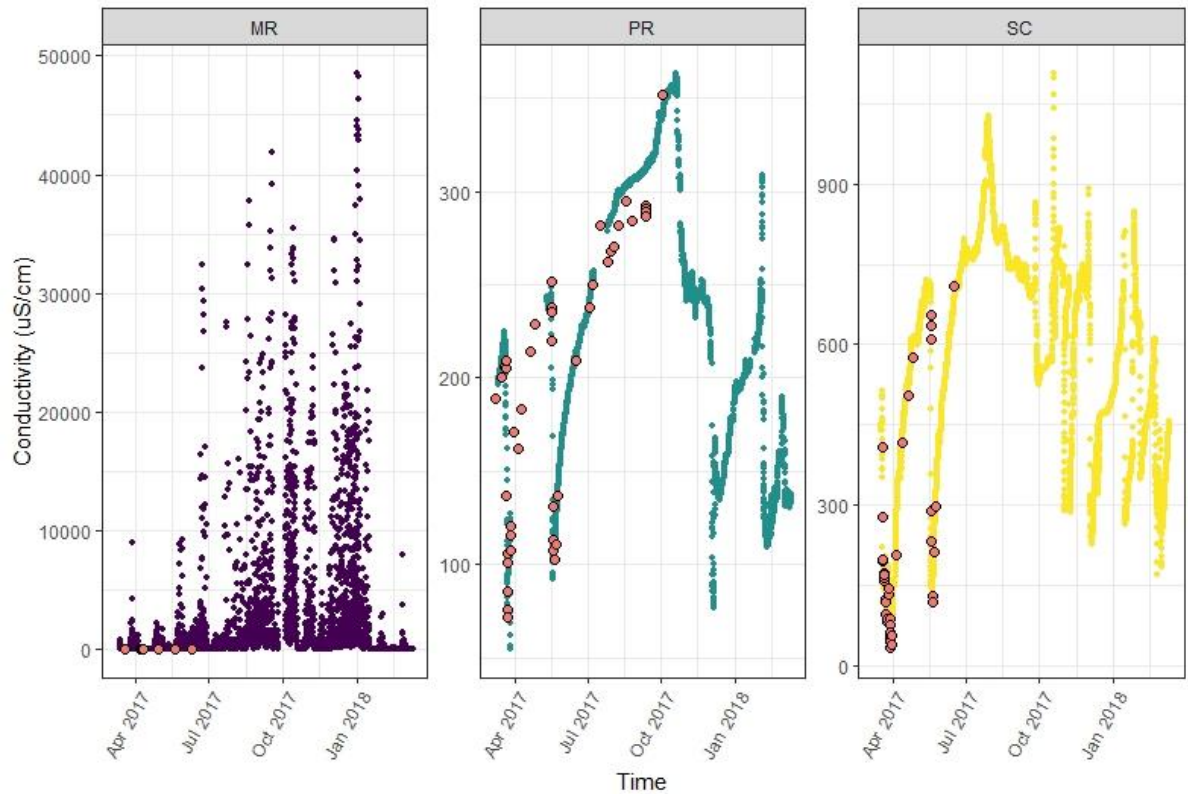

**Figure 2.** Laboratory-measured (large, outlined points) and *in situ* sensor-measured conductivity (μS/cm) between March 2017 and March 2018 at Mulgrave River (MR; purple points), Pioneer River (PR, green points) and Sandy Creek (SC; yellow points). Gaps in time series are missing observations.

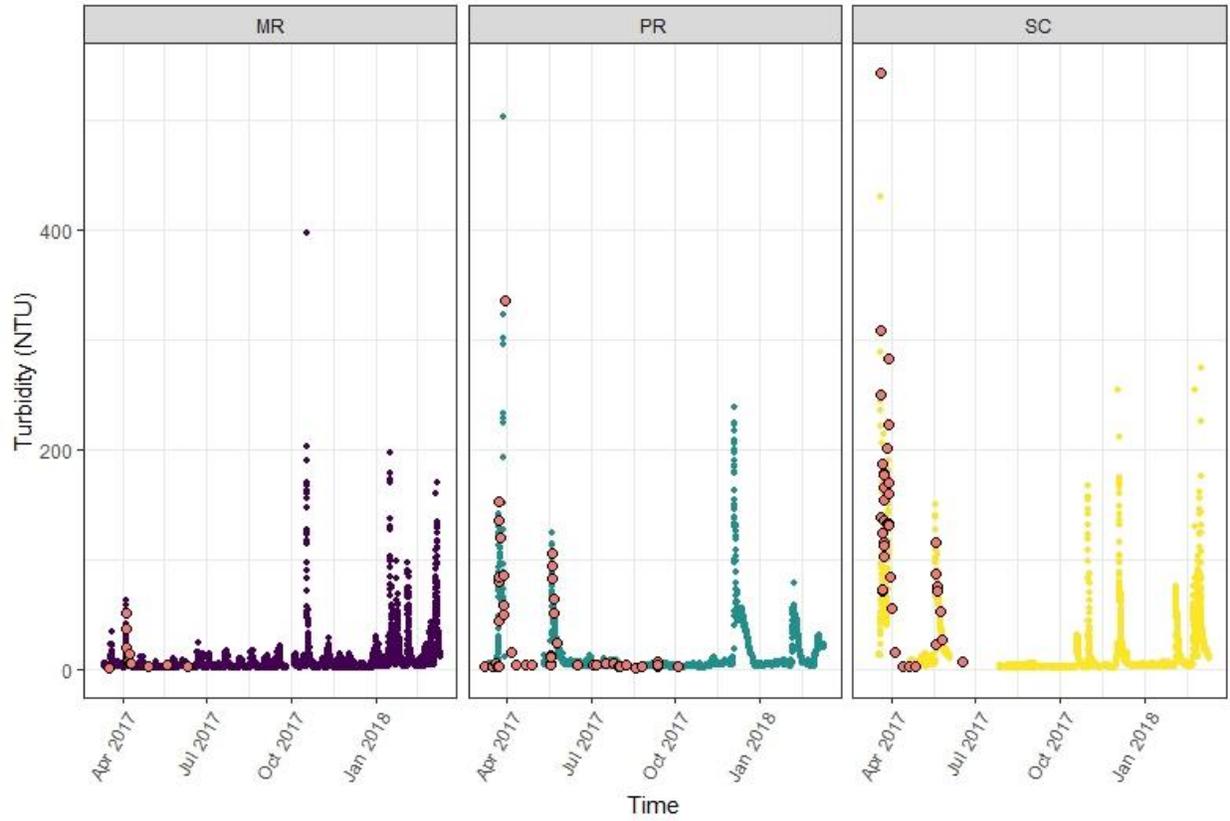

**Figure 3.** Laboratory-measured (large, outlined points) and *in situ* sensor-measured turbidity (NTU) between March 2017 and March 2018 at Mulgrave River (MR; purple points), Pioneer River (PR, green points) and Sandy Creek (SC; yellow points). Gaps in time series are missing observations.

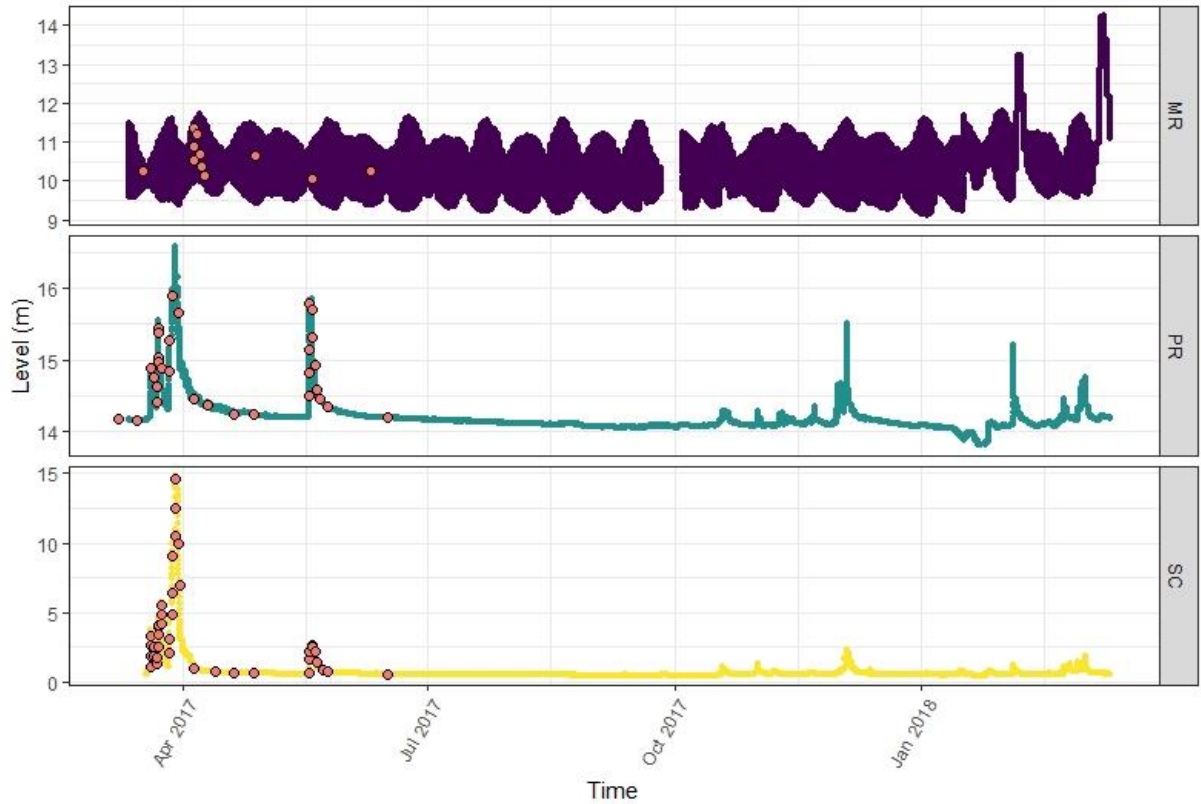

**Figure 4.** River level (m) measured on-site at the time of water sample collection (large, outlined points) and by *in situ* sensors between March 2017 and March 2018 at Mulgrave River (MR; purple points), Pioneer River (PR, green points) and Sandy Creek (SC; yellow points). Gaps in time series are missing observations.

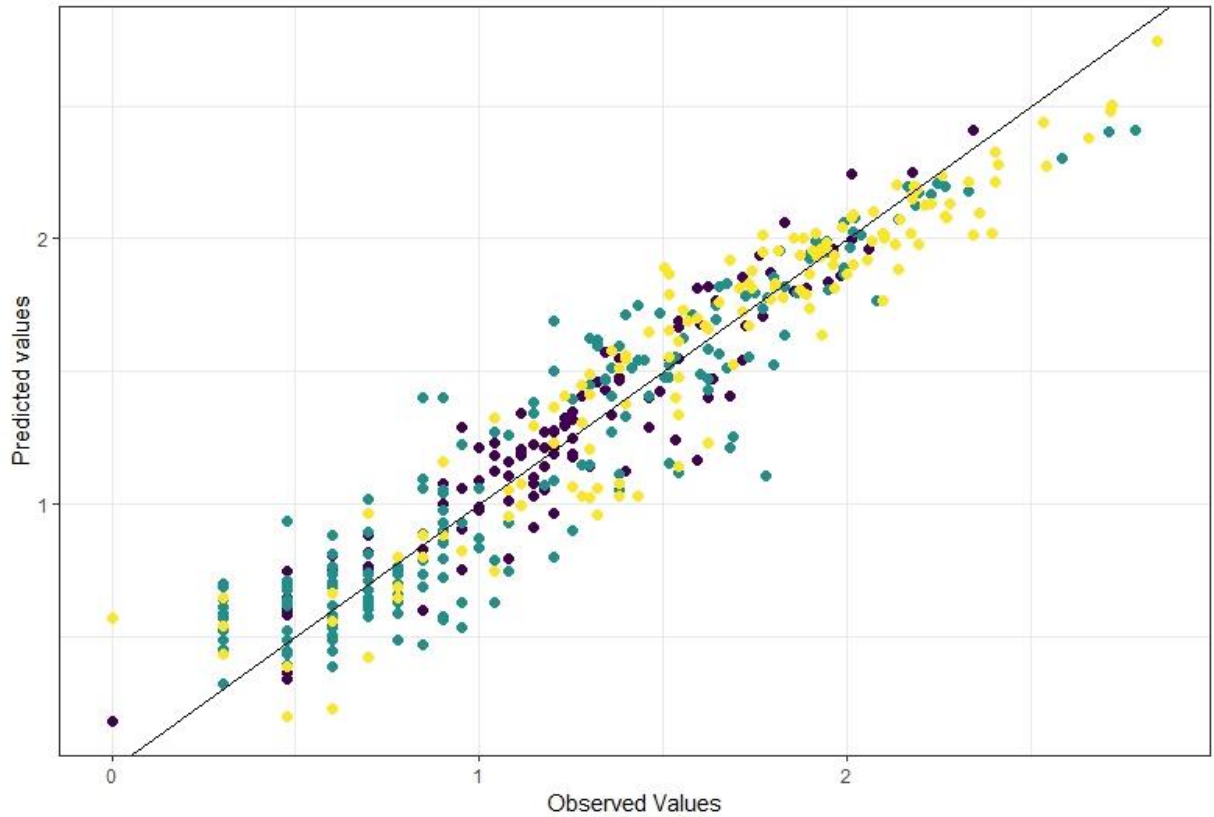

**Figure 5.** Observed versus cross-validated prediction values of $\log_{10}$-transformed total suspended solids (TSS, mg, L) from the final TSS model, with data from each site shown in purple (Mulgrave River, MR), green (Pioneer River, PR) and yellow (Sandy Creek, SC). The black line shows the 1:1 ratio.

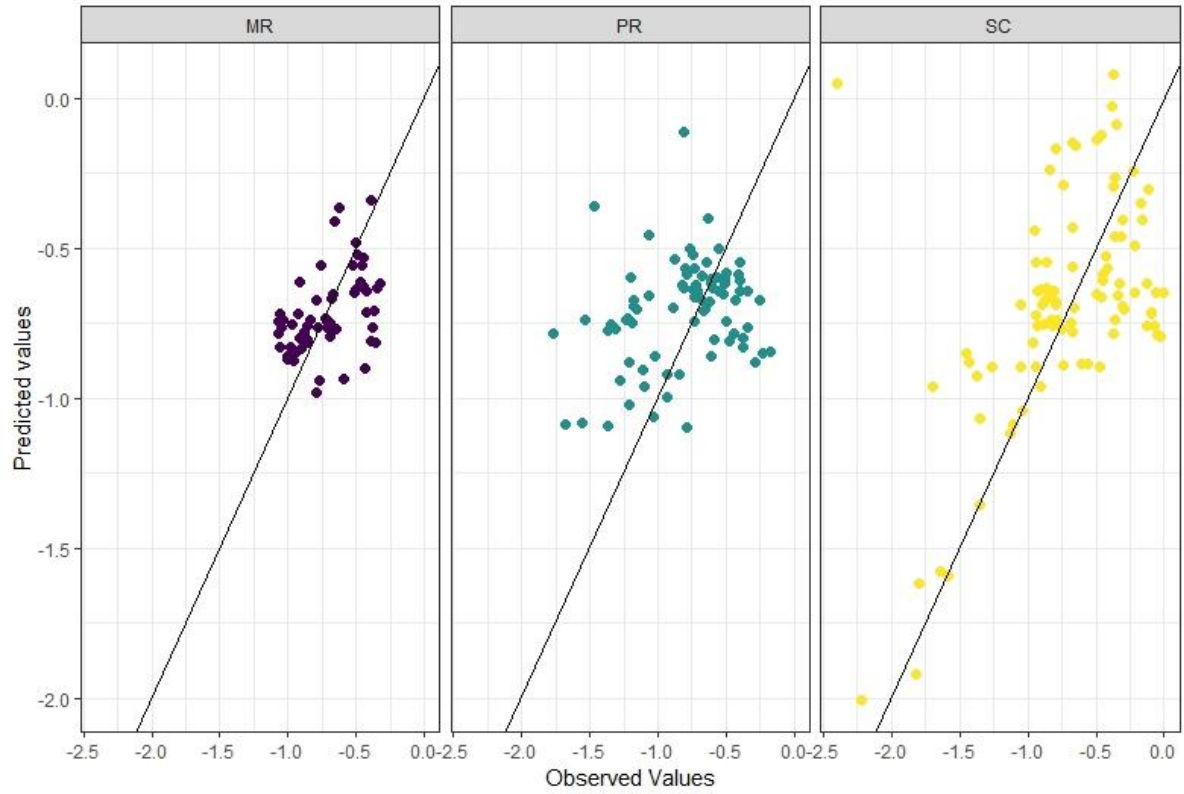

**Figure 6.** Observed versus cross-validated prediction values of log$_{10}$-transformed oxidized nitrogen (NOx; mg/L) from the final composite NOx model for each site (Mulgrave River, MR; purple points; Pioneer River, PR; green points; Sandy Creek, SC; yellow points). Black lines show the 1:1 ratios.

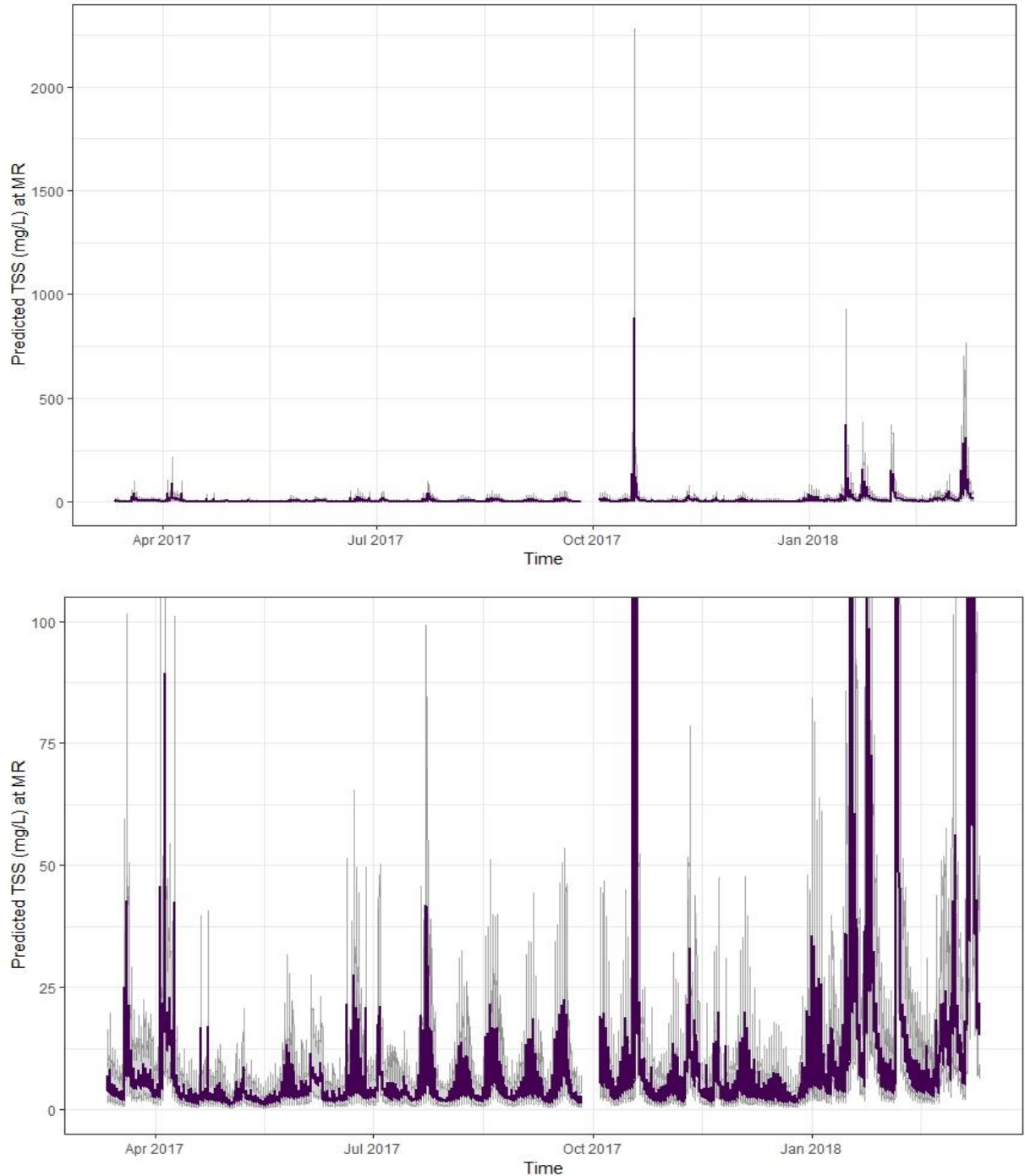

**Figure 7.** Concentration of total suspended solids (TSS, mg/L) at Mulgrave River (MR) predicted using the final TSS model from *in situ* sensor turbidity data measured between March 2017 and March 2018. Gray lines show upper and lower boundaries of the 95% prediction interval, and the purple line the predicted TSS values through time. The lower plot contains the same data as the upper plot, but has a narrower scale along the y-axis to show detail. Gaps indicate periods of missing data in the sensor time series.

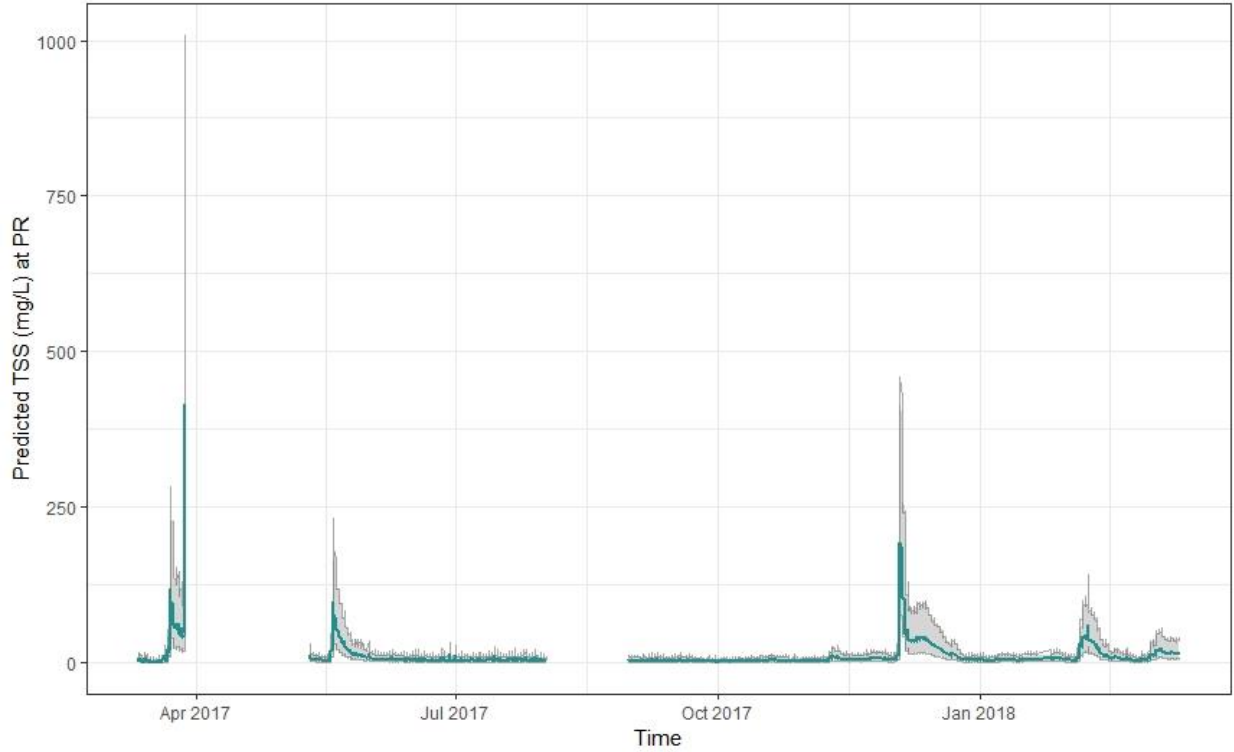

**Figure 8.** Concentration of total suspended solids (TSS, mg/L) at Pioneer River (PR) predicted using the final TSS model from *in situ* sensor turbidity data measured between March 2017 and March 2018. Gray lines show upper and lower boundaries of the 95% prediction interval, and the green line the predicted TSS values through time. Gaps indicate periods of missing data in the sensor time series.

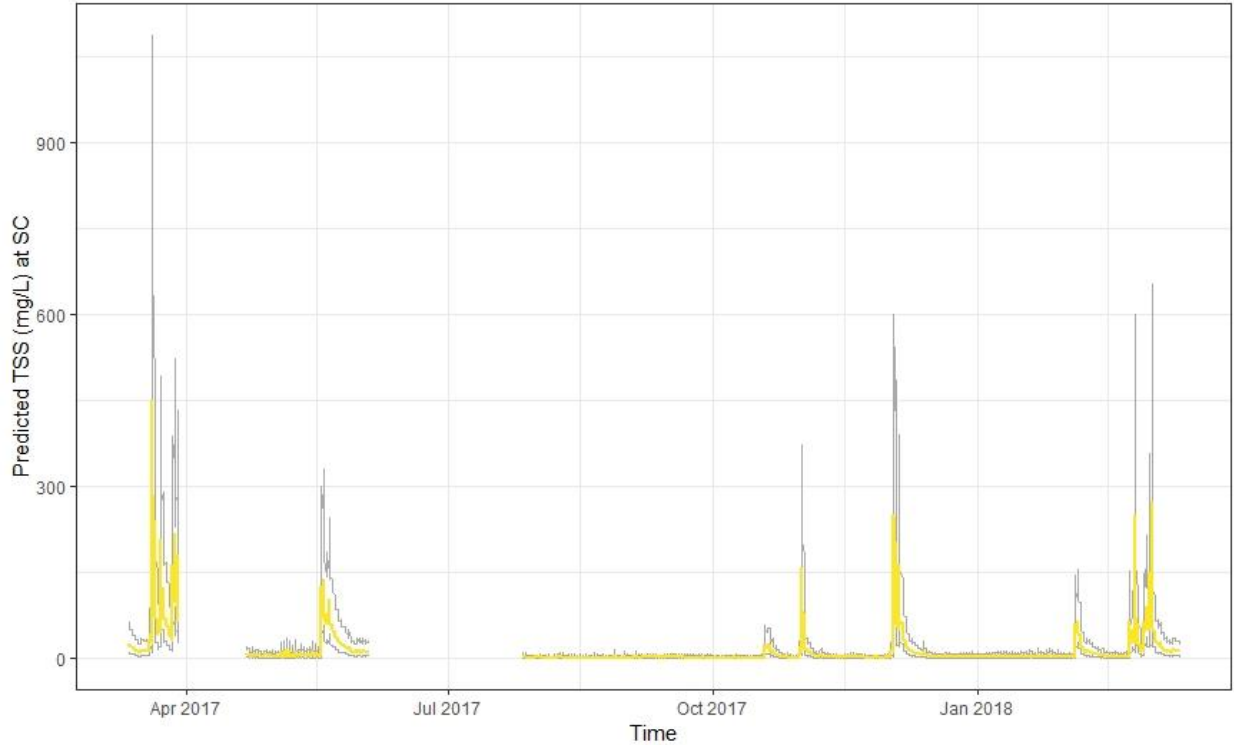

**Figure 9.** Concentration of total suspended solids (TSS, mg/L) at Sandy Creek (SC) predicted using the final TSS model from *in situ* sensor turbidity data measured between March 2017 and March 2018. Gray lines show upper and lower boundaries of the 95% prediction interval, and the yellow line the predicted TSS values through time. Gaps indicate periods of missing data in the sensor time series.

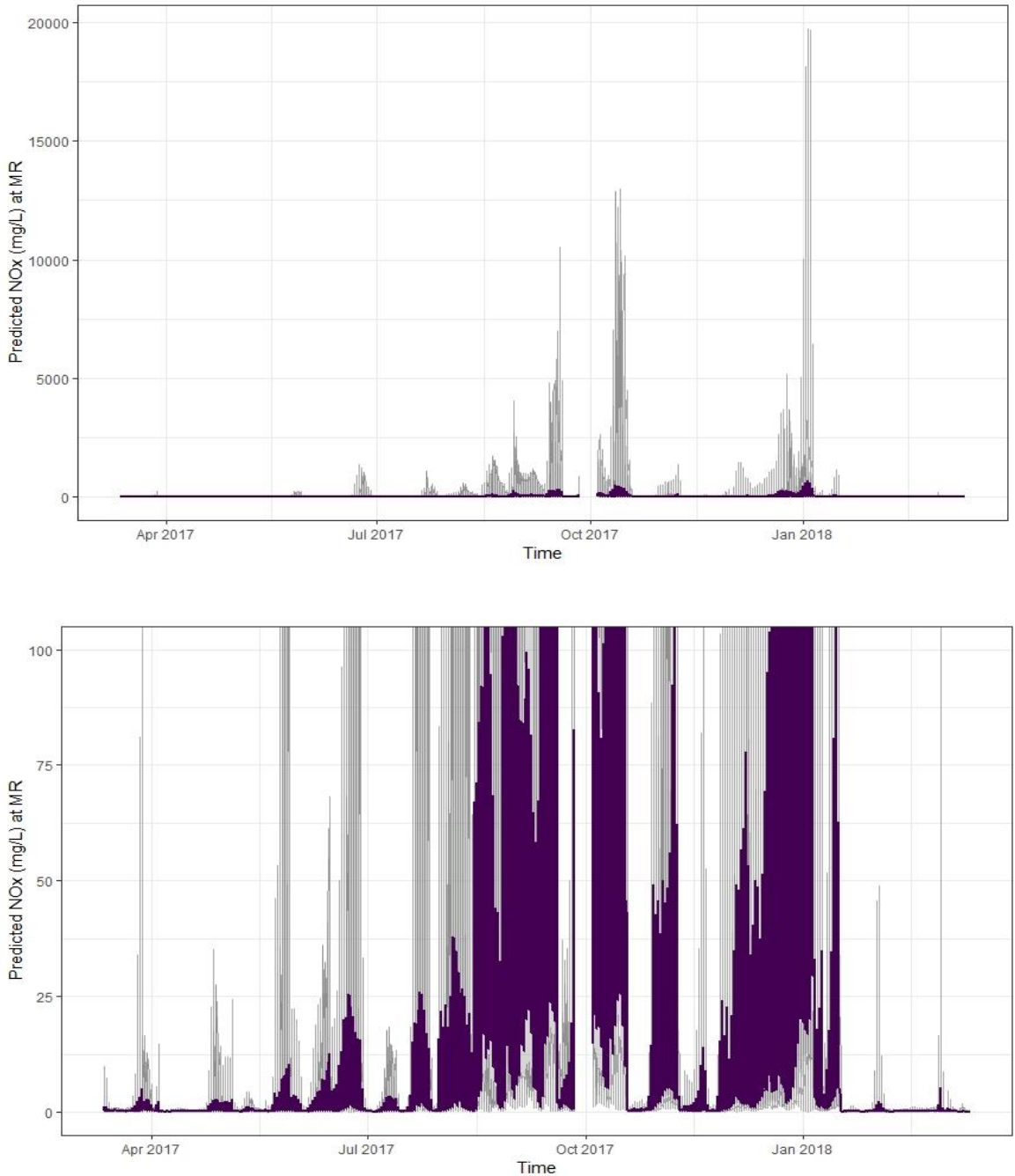

**Figure 10.** Concentration of oxidized nitrogen (NOx, mg/L) at Mulgrave River (MR) predicted using the final NOx model from *in situ* sensor turbidity, conductivity and level data measured between March 2017 and March 2018. Gray lines show upper and lower boundaries of the 95% prediction interval, and the purple line the predicted NOx values through time The lower plot contains the same data as the upper plot, but has a narrower scale along the y-axis to show detail. Gaps indicate periods of missing data in the sensor time series.

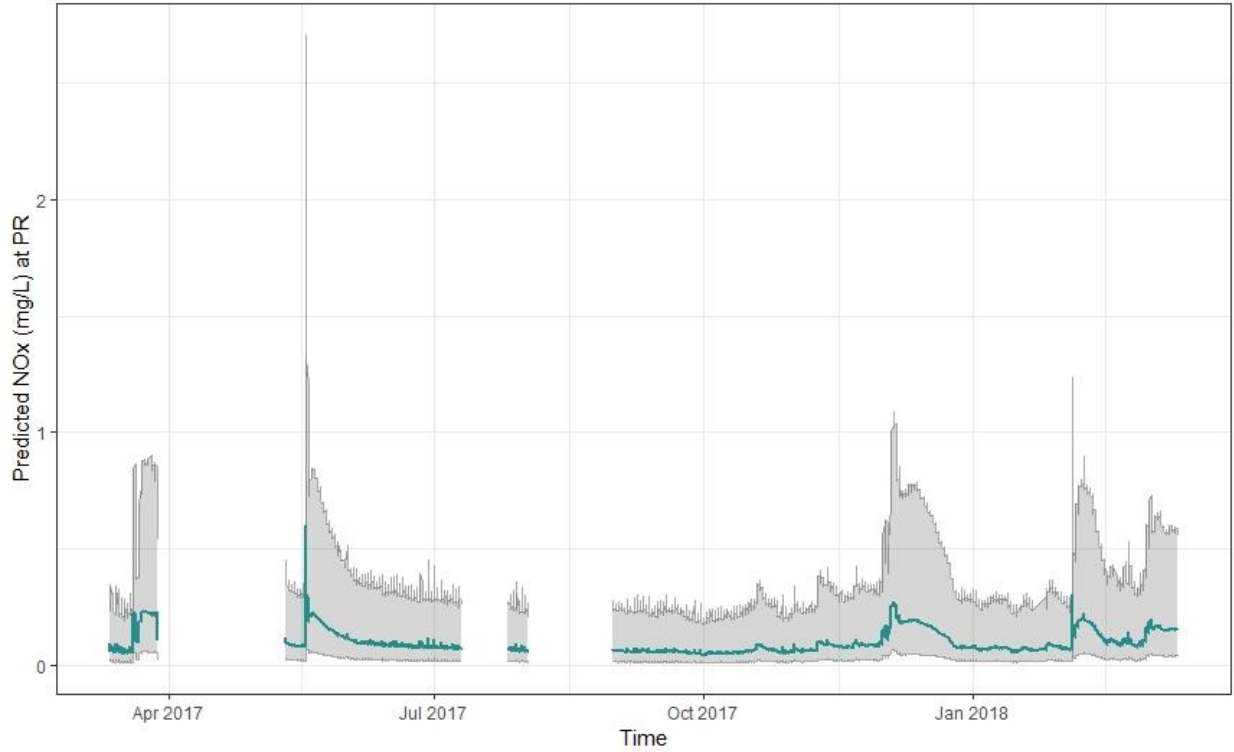

**Figure 11.** Concentration of oxidized nitrogen (NOx, mg/L) at Pioneer River (PR) predicted using the final NOx model from *in situ* sensor turbidity, conductivity and level data measured between March 2017 and March 2018. Gray lines show upper and lower boundaries of the 95% prediction interval, and the green line the predicted NOx values through time. Gaps indicate periods of missing data in the sensor time series.

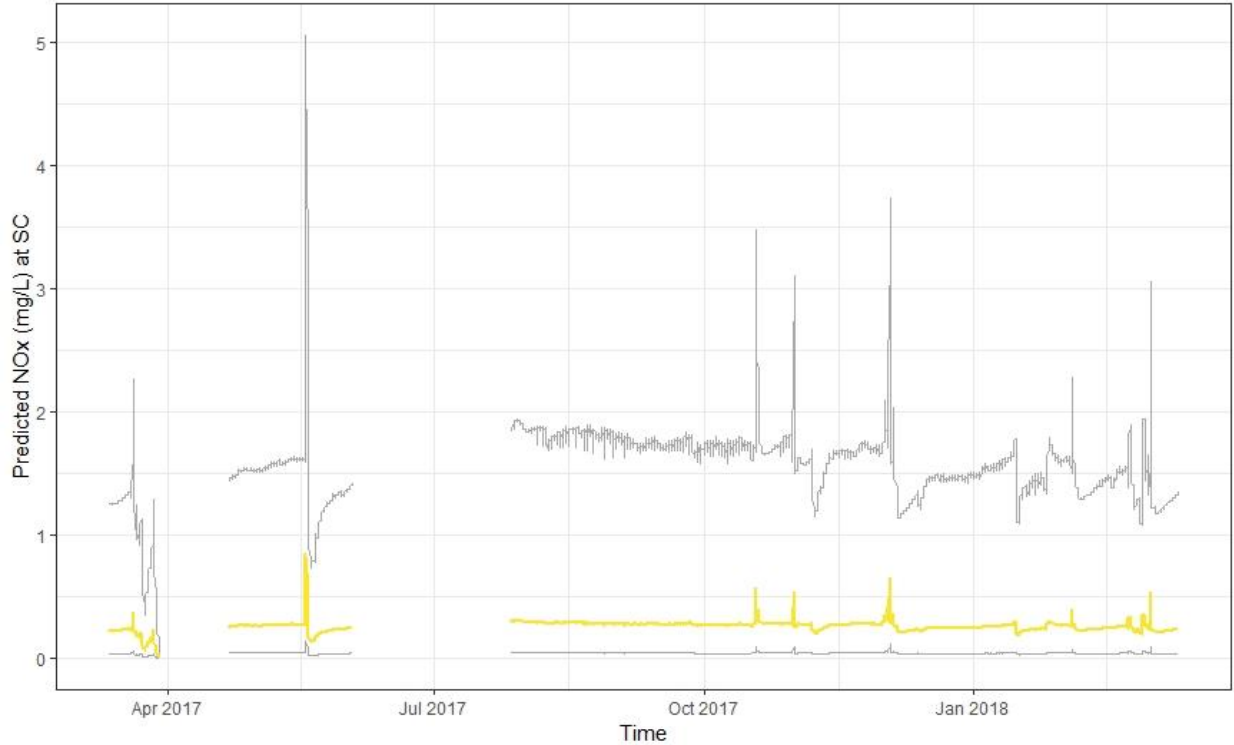

**Figure 12.** Concentration of oxidized nitrogen (NOx, mg/L) at Sandy Creek (SC) predicted using the final NOx model from *in situ* sensor turbidity, conductivity and level data measured between March 2017 and March 2018. Gray lines show upper and lower boundaries of the 95% prediction interval, and the yellow line the predicted NOx values through time. Gaps indicate periods of missing data in the sensor time series.

**Tables**

**Table 1.** The range (minimum to maximum) and number of observations (shown in parentheses) for total suspended solids (TSS), oxidized nitrogen (NOx), turbidity, conductivity, and river level as measured in the laboratory or *in situ* by automated sensors, by site.

| Measure (unit) | Source | Mulgrave River | Pioneer River | Sandy Creek |
|---|---|---|---|---|
| TSS (mg/L) | Laboratory | 1.0 - 221 (106) | 2.0 - 609 (193) | 1.0 - 700 (143) |
| NOx (mg/L) | Laboratory | 0.015 - 0.477 (106) | 0.002 - 1.000 (183) | 0.004 - 0.986 (113) |
| Turbidity (NTU) | Laboratory | 1.2 - 143 (106) | 1.3 - 335 (193) | 1.6 - 542 (143) |
| | Sensor | 0.58 - 396 (6276) | 0.61 - 503 (6158) | 1.15 - 430 (5399) |
| Conductivity (µS/cm) | Laboratory | 30 - 2900 (106) | 52 - 352 (183) | 33 - 988 (113) |
| | Sensor | 11.57 - 48453 (6278) | 54.41 - 363 (6163) | 29.68 - 1105 (5401) |
| Level (m) | Laboratory* | 9.94 - 11.70 (58) | 13.98 - 15.89 (81) | 0.46 - 14.55 (94) |
| Measure (unit) | Sensor | 9.12 - 14.24 (37 085) | 13.81 - 16.58 (9893) | 0.46 - 14.77 (5401) |

*Recorded on-site at the time of grab-sample collection.

**Table 2.** Cross validation (*cv*) statistics for the final total suspended solids (TSS) and oxidized nitrogen (NOx) models (r-squared, $R^2$; root mean square error, *RMSE*; prediction coverage, *PC*) and the correlation parameter ($\phi$).

| Model | $cvR^2$ | cvRMSE (back-transformed equivalent in mg/L) | cvPC | $\phi$ (95% confidence interval) |
|---|---|---|---|---|
| TSS | 90.74% | 0.1799 (29.62) | 100% | 0.8737 (0.8304 - 0.9072) |
| NOx (MR) | 25.81% | 0.1988 (0.0987) | 74.14% | 0.8612 (0.7492 - 0.9280) |
| NOx (PR) | 17.42% | 0.3315 (0.1553) | 67.90% | 0.8607 (0.7306 - 0.9337) |
| NOx (SC) | 35.05% | 0.3965 (0.2866) | 95.75% | 0.8657 (0.7245 - 0.9405) |

**Table 3.** ANOVA table for the fixed effects in the final total suspended solids (TSS) model.

| Coefficient | Degrees of freedom | *F*-value | *p*-value |
|---|---|---|---|
| Intercept | 1, 434 | 11241 | <0.0001 |
| Turbidity | 1, 434 | 2449 | <0.0001 |
| Site | 2, 434 | 8.2 | 0.0003 |
| T15 | 1, 434 | 19.2 | < 0.0001 |
| Turbidity × Site | 2, 434 | 8.2 | 0.0003 |
| Turbidity × T15 | 1, 434 | 8.7 | 0.0034 |

Note: Turbidity was $\log_{10}$-transformed. ANOVA was applied on the fitted model, which accounted for the continuous AR(1) error structure.

**Table 4.** ANOVA table for the fixed effects in the final oxidized nitrogen (NOx) model for each site.

| Site | Coefficient | Degrees of freedom | $F$-value | $p$-value |
|---|---|---|---|---|
| MR | Intercept | 1, 52 | 820.6 | <0.0001 |
|  | Conductivity | 1, 52 | 39.1 | <0.0001 |
|  | Turbidity | 1, 52 | 0.3 | 0.6147 |
|  | Level | 1, 52 | 2.0 | 0.1618 |
|  | Conductivity × Level | 1, 52 | 2.3 | 0.1367 |
|  | Turbidity × Level | 1, 52 | 0.02 | 0.8759 |
| PR | Intercept | 1, 75 | 568.6 | <0.0001 |
|  | Conductivity | 1, 75 | 15.4 | 0.0002 |
|  | Turbidity | 1, 75 | 0.03 | 0.8592 |
|  | Level | 1, 75 | 9.8 | 0.0024 |
|  | Conductivity × Level | 1, 75 | 5.8 | 0.0183 |
|  | Turbidity × Level | 1, 75 | 9.9 | 0.0024 |
| SC | Intercept | 1, 88 | 283.4 | <0.0001 |
|  | Conductivity | 1, 88 | 44.8 | <0.0001 |
|  | Turbidity | 1, 88 | 9.9 | 0.0022 |
|  | Level | 1, 88 | 0.9 | 0.3428 |
|  | Conductivity × Level | 1, 88 | 6.4 | 0.0134 |
|  | Turbidity × Level | 1, 88 | 3.4 | 0.0688 |

Note: Conductivity, turbidity and level were $\log_{10}$-transformed. ANOVA was applied on the fitted model, which accounted for the continuous AR(1) error structure.